\documentclass[12pt]{article}
\usepackage{hyperref}
\usepackage[utf8x]{inputenc}
\usepackage{extarrows}
\usepackage{enumerate}
\usepackage{tikz-cd}

\usepackage[margin=3cm]{geometry}
\usepackage{amsmath,amssymb}
\usepackage[mathscr]{euscript}
\usepackage{color}
\usepackage{pstricks}
\usepackage{graphicx}
\usepackage[all]{xy}
\usepackage{tabularx}
\usepackage{bbm}

\newtheorem{defi}{Definition}[section]
\newtheorem{lemm}{Lemma}[section]
\newtheorem{thm}{Theorem}[section]
\newtheorem{coro}{Corollary}[section]

\newcommand{\bbox}{\normalsize {}%
        \nolinebreak \hfill $\blacksquare$ \medbreak \par}
\newenvironment{proof}{\noindent\emph{Proof} ---}{\bbox\vspace{0,15cm}}
\newcommand{\wh}[1]{\widehat{#1}}
\newcommand{\dd}{\mathrm{d}}
\newcommand{\ii}{\mathrm{i}}
\newcommand{\me}{\textrm{e}}
\newcommand{\ol}[1]{\overline{#1}}
\newcommand{\lto}{\longrightarrow}

\allowdisplaybreaks

\newcommand{\mm}[1]{\mathrm{#1}}
\newcommand{\mb}[1]{\mathbb{#1}}

\newcommand{\mn}[1]{\mathbf{#1}}

\DeclareMathOperator{\dom}{Dom}

\newcommand{\fk}[1]{\mathfrak{#1}}

\newcommand{\tts}{\textstyle}

\newcommand{\bank}[1]{\big\langle #1 \big\rangle}

\newcommand{\sank}[1]{\langle #1 \rangle}

\DeclareMathOperator{\ttr}{tr}

\DeclareMathOperator{\vol}{vol}
\DeclareMathOperator{\ddiv}{div}

\DeclareMathOperator{\detz}{det_{\zeta}}

\newcommand{\nrm}[1]{\left\lVert#1\right\rVert}

\newcommand{\one}{\mathbbm{1}}

\renewcommand{\tilde}{\widetilde}
\renewcommand{\ge}{\geqslant}
\renewcommand{\le}{\leqslant}
\newcommand{\defeq}{\overset{\mathrm{def}}{=}}

\title{Asymptotics of zeta determinants of Laplacians\\ on large degree abelian covers.}
\author{ Nguyen Viet Dang, Jiasheng Lin, Frédéric Naud  }
\date{}
\begin{document}
\maketitle

\begin{abstract}
Let $(M,g)$ be some smooth, closed, compact Riemannian manifold and 
$(M_N\lto M)_N$ be an increasing sequence of large degree cyclic covers of $M$ that converges when $N\rightarrow +\infty$, in a suitable sense, to some limit $\mathbb{Z}^p$ cover $M_\infty$ over $M$. Motivated by recent works on zeta determinants on random surfaces, Kirchhoff formula in graph theory, and some natural questions in Euclidean quantum field theory,  we show the convergence of the sequence $ \frac{\log\det_\zeta(\Delta_{N})}{\text{Vol}(M_N)} $ when $N\rightarrow +\infty$ where $\Delta_N$ is the Laplace-Beltrami operator on $M_N$.  We also generalize our results to the case of twisted Laplacians coming from certain flat unitary vector bundles over $M$.
%
\end{abstract}

\tableofcontents

\section{Introduction}
\label{s:abeliancovers}

In statistical physics, for $m\geqslant 0$,
 the discrete Gaussian Free Field (DGFF) is some ferromagnetic spin system which is defined on a discrete box $\Lambda\subset \mathbb{Z}^d$ of finite size. 
 The configuration space of the DGFF is the space $\mathbb{R}^\Lambda$
weighted with the probability measure 
$ \frac{1}{Z_\Lambda}\exp\left( -S(\sigma) \right) d^\Lambda\sigma $ where $S(\sigma):=\sum_{i\sim j\in \Lambda^2} \vert \sigma_i-\sigma_j\vert^2 -\sum_{i\in \Lambda} m^2\sigma_i^2$ is the discrete Dirichlet functional
and $Z_\Lambda$ is the partition function given by $ Z_\Lambda:= \int_{\sigma \in \mathbb{R}^\Lambda}  \exp\left( -S(\sigma) \right) d^\Lambda\sigma$. 
In the discrete setting, the Dirichlet action functional $S(\sigma)$ is a quadratic form on $\mathbb{R}^\Lambda$ which rewrites  $S(\sigma)=\sum_{i\in \Lambda} \sigma_i(\Delta_\Lambda \sigma)_i+m^2\sigma_i^2 $ where $\Delta_\Lambda$ is the discrete Laplacian and therefore, the partition function $Z_\Lambda$ of the discrete GFF is a simple Gaussian integral which is easily expressed in terms of $\Delta_\Lambda$ as~:
\begin{align*}
 Z_\Lambda= \det(\Delta_\Lambda+m^2)^{-\frac{1}{2}} . 
\end{align*}

Assume now our spin system is defined on some smooth compact manifold
$M$ and we would like to define a certain continuum limit of the model. It is expected that a suitable scaling limit of the DGFF is described by a certain fundamental model from Euclidean Quantum Field Theory, the Gaussian Free Field (GFF). The partition function of the massive Gaussian free field is formally denoted by
\begin{align*}
Z_M=\int_{\varphi\in \text{Functions on }M} \exp\left( -\int_M \vert \nabla \varphi \vert^2 + m^2\varphi^2 dV_g \right) d\varphi.    
\end{align*}
On the manifold $M$, since the space of functions is infinite dimensional and the Laplace Beltrami operator is no longer a finite dimensional matrix, one has to regularize the  determinant or equivalently one has to regularize the infinite product $\prod_{\lambda\in \sigma(\Delta)\setminus \{0\}} (\lambda+m^2)$.
Inspired by the seminal work of Ray--Singer~\cite{RaSi} and also by the more physically oriented paper
of Hawking~\cite{Hawking},
a common choice in theoretical physics is to use zeta regularization to give a mathematical meaning to free fields partition functions. So $Z_M$ is defined to be 
\begin{align*}
Z_M:=\text{det}_\zeta\left(\Delta_g+m^2 \right)^{-\frac{1}{2}}    
\end{align*}
where we refer to paragraph~\ref{sss:statement} for a precise definition of 
zeta determinants.

From the point of view of statistical physics, 
the free energy $\log Z_M$ is expected to grow as the size of the system, here the natural system being our Riemannian manifold $M$. So the natural \textbf{intensive quantity} should be the 
density of free energy defined as~\cite[eq (24.28) p.~144]{Schwarz}
\begin{equation}\label{eq:freeenergy}
 F= \frac{\log (Z_M)}{\text{Vol}(M)} .
\end{equation}

It is expected, as the volume of our Riemannian manifold gets larger in a certain limiting sense, that 
the above quantity converges as the volume of $M$ goes to infinity. 
It means there exists a \textbf{density of free energy in the thermodynamic limit}, which is also interpreted as the energy density of the ground state of a quantum system described by the GFF. 
Inspired by the recent results of the third author~\cite{Na23} as well as several studies on the spectral theory of Laplacians on random surfaces~\cite{MonkMarklov} and motivated by problems of taking large volume limits in
quantum field theory~\cite{Lin24},
the goal of the present paper is to establish the existence of a limit for the quantity $F$ defined by equation~(\ref{eq:freeenergy}) for suitable sequences of cyclic covers of large degree that \emph{converge} in some appropriate sense to some $\mathbb{Z}^p$-periodic Riemannian manifolds obtained as abelian cover of some initial closed compact Riemannian manifold $M$.

\subsection{Mathematical setting.}
\label{sss:setting}
The aim of the present paragraph is to describe our mathematical setting.
Let $(M,g)$ denotes a smooth, closed, compact Riemannian manifold of dimension $d$. 
We shall start by giving some way to manufacture towers of abelian covers of $M$ following~\cite{JaNaSo}.
Consider any group homomorphism of the form
$$\rho: \pi_1(M)\lto H_1(M,\mathbb{Z})\lto \mathbb{Z}$$
where the last arrow $H_1(M,\mathbb{Z})\lto \mathbb{Z}$ has infinite elements in its image
\footnote{This assumption is important: it is always satisfied for compact orientable surfaces, but there exist many orientable 3-manifolds with no cyclic covers.
To give an explicit example, the $5$-surgery on a figure eight knot complement yields a hyperbolic $3$-manifold whose first homology group is 
$\mathbb{Z}/5\mathbb{Z}$, see \cite{Rolfsen}, chapter 9.} 
Then $\ker(\rho)$ is a subgroup of the fundamental group
$\pi_1(M)$. We denote by $\widehat{M}$
the universal cover of $M$, $\ker(\rho)$ is a subgroup of the deck group of $\widehat{M}\lto M$ and the quotient 
$M_\infty:= \widehat{M}/\ker(\rho) $
is an abelian cover of $M$ with deck group $\mathbb{Z}$, in fact the quotient
$  \widehat{M}/\ker(\rho) $ is a periodic Riemannian manifold endowed with an 
isometric action of $\mathbb{Z}$.
We denote by $\rho_N:\pi_1(M)\lto \mathbb{Z}\lto \mathbb{Z}/N\mathbb{Z}$
the composition of $\rho$ with mod $N$ reduction.
Then the quotient $M_N:= \widehat{M}/\ker(\rho_N)$ is an abelian cover of $M$ with deck group $\mathbb{Z}/N\mathbb{Z}$ and when $N\rightarrow +\infty$, we can think of the sequence $(M_N)_N$ as an increasing sequence of abelian covers that \emph{converges} to the periodic cover $M_\infty$. Denoting by $\gamma$ the generator of the $\mathbb{Z}$-action on $M_\infty$, one can also think of $M_N$ as the quotient $M_\infty/\left\langle \gamma^N \right\rangle$ where $\left\langle \gamma^N \right\rangle$ denotes the subgroup of isometries generated by $\gamma^N$.

\subsection{Statement of the main result.}
\label{sss:statement}

We keep the notations from the previous paragraph, denote by $\Delta_N$ (resp $\Delta_\infty$) the Laplace-Beltrami operator on $M_N$ (resp $M_\infty$), we can state the main results of our note.
%
We consider $M_N\lto M$ a sequence of cyclic cover of degree $N$ as defined in paragraph~\ref{sss:setting}, we denote by $\det_\zeta(\Delta_{N})$ the zeta regularized determinant of $\Delta_{N}$ defined
as~\cite{RaSi}
\begin{eqnarray}
\text{det}_\zeta(\Delta_{M_N})=\exp(-\zeta^\prime(0)),\,\quad \quad \zeta_{\Delta_N+1}(s)=\sum_{\lambda\in\sigma(\Delta_N)\setminus \{0\}}\lambda^{-s}
\end{eqnarray}
where the spectral zeta function $\zeta_{\Delta_N}(s)
$ has holomorphic continuation near $s=0$ (note that we omitted the prime, which is a more common notation, indicating exclusion of zero eigenvalue).
 We show in the present note the following:
\begin{thm}\label{thm:convergencewithmass}
Under the above assumptions, we find that the sequence~:
\begin{eqnarray}
\boxed{\frac{\log\left(\det_\zeta\left( \Delta_{M_N}\right) \right)}{\vol(M_N)}}
\end{eqnarray}
has a limit when $N\rightarrow +\infty$ which is expressed in terms of the heat kernel of $M_\infty$.
\end{thm}

We mention here that this limit is in general non-vanishing, as demontrated by the example of flat tori, see below. It would be interesting to know more about the possible limit points of $\log\left(\det_\zeta\left( \Delta_{M_N}\right) \right)/\text{Vol}(M_N)$, in particular can it fill an interval ? We will see examples below where this limit can be negative. Does the sign of the limit reflect some of the geometry of manifolds $M_N$ ?

In subsection~\ref{s:bundlecase}, we indicate how to generalize the above Theorem to certain twisted Laplacians of the form $\nabla^*\nabla$ where $\nabla$ is a flat unitary connection. The generalization relies on simple diamagnetic inequalities comparing 
 heat kernels of magnetic Laplacians and scalar heat kernels.

\subsection{Some examples}

In this paragraph, we shall motivate our analysis by discussing discrete known analogues of our result on graphs,
as well as the examples of flat tori and hyperbolic surfaces.
\subsubsection{Combinatorial Laplacians on regular graphs}
Let $X$ be a connected finite graph with $|X|$ vertices. The \textsf{graph Laplacian} $\Delta_X$ on $X$ is defined by
\begin{equation}
  (\Delta_X f)(v)\defeq d(v)f(v)-\sum_{w\sim v}f(w)
  \label{}
\end{equation}
where~$d(v)$ is the \textit{degree} of the vertex~$v$ and~$w\sim v$ means $w$ is joined to $v$ by an edge of $X$. Now $\Delta_X$ is a finite non-negative symmetric matrix and $0$ is a simple eigenvalue. We consider here
\begin{equation}
  {\tts\det'}(\Delta_X)\defeq\prod_{\lambda\in \sigma(\Delta_X)\setminus\{0\}} \lambda,
  \label{}
\end{equation}
and we are interested in the existence of possible limits
\begin{equation}
  \lim_{\text{``}|X|\to \infty\text{''}}\frac{\log{\tts\det'}(\Delta_X)}{|X|}
  \label{}
\end{equation}
in various senses.
In this case a useful tool is the famous \textit{matrix tree theorem} of Kirchhoff which says
\begin{thm}
  [Kirchhoff, \cite{Sarnak-moser}, \cite{Kirchhoff}] 
 $
    {\tts\det'}(\Delta_X)=|X|K(X)
  $.
\end{thm}
Here $K(X)$ denotes the number of \textsf{spanning trees} in~$X$, that is, trees (connected subgraphs with no cycles) containing all vertices. Therefore, if limits do exist, we have
\begin{equation}
  \lim_{\text{``}|X|\to \infty\text{''}}\frac{\log{\tts\det'}(\Delta_X)}{|X|}=\lim_{\text{``}|X|\to \infty\text{''}}\frac{\log K(X)}{|X|}.
  \label{}
\end{equation}
Moreover, if $\mathcal{G}_{n,k}$ denotes the set of $k$-regular graphs with size $n$, a classical result due to McKay  \cite{McKay} says that for $k\ge 3$
$$\limsup_{n\to\infty, X_{k,n}\in \mathcal{G}_{n,k}} \frac{\log K(X_{k,n})}{n}=\log c_k $$
with
\begin{equation}
  c_k\defeq \frac{(k-1)^{k-1}}{(k^2-2k)^{k/2-1}}.
  \label{eqn-mckay-law-const}
\end{equation}
See e.g.\ the introductions of \cite{GKW14}, \cite{Rosen-Tenen-spanning} for more information. This limsup is actually a limit for typical random graphs and sequences of Ramanujan graphs.

\bigskip
On the other hand, in the simplest case which is an exact analogue of the geometric set-up considered in this paper, we look at the cyclic graph~$C_n$ (vertices $=$ elements of the cyclic group $\mb{Z}_n$, $[j]$ joined with $[j+1]$). In $C_n$ a spanning tree is simply a subgraph with one edge omitted from $C_n$. Hence~$K(C_n)=n$ and
\begin{equation}
  \lim_{n\to \infty}\frac{\log {\tts\det'}(\Delta_{C_n})}{|C_n|}=\lim_{n\to\infty}\frac{\log(n^2)}{n}=0.
  \label{}
\end{equation}

Consider now the ``2-dimensional'' version of $C_n$, the square lattice on the 2-torus with vertex set $\mb{Z}_n\times\mb{Z}_n$, or more precisely the (uncolored, undirected) \textit{Cayley graph} $\Gamma(\mb{Z}_n\times\mb{Z}_n,\mathcal{E})$ where
\begin{equation}
  \mathcal{E}=\{([1],0),(0,[1]),([n-1],0),(0,[n-1])\}.
  \label{}
\end{equation}
Then it is possible to show that \cite{Lyonsspanning}
\begin{equation}
  \lim_{n\to\infty}\frac{\log{\tts\det'}(\Delta_{\Gamma(\mb{Z}_n^2,\mathcal{E})})}{n^2}=\alpha
  \label{}
\end{equation}
with~$0<\alpha<\log c_4$ where~$c_4$ is a known constant defined in (\ref{eqn-mckay-law-const}) below. More generally, for cayley graphs of non-cyclic abelian groups, the above limit always exists and is $>0$ but below McKay's constant $\log(c_k)$.

\subsubsection{Determinants of Laplacians on flat Tori}

Consider the torus~$\mb{C}/\Lambda_{\tau}$ where~$\Lambda_{\tau}=\mb{Z}\oplus \tau\mb{Z}$ with the modular parameter~$\tau$ in the upper-half plane 
$$\mathbb{H}:=\{\fk{Im}(\tau)>0\}.$$
Then it is well-known that~$\detz(\Delta_{\mb{C}/\Lambda_{\tau}})$ has the expression~\cite{Kierlanczyk}, \cite{Sarnak-moser} 
\begin{equation}
  \detz(\Delta_{\mb{C}/\Lambda_{\tau}})=(\fk{Im}\, \tau)^2|\eta(\tau)|^4
  \label{eqn-formula-detz-torus}
\end{equation}
where~$\eta(\tau)$ is the Dedekind eta modular form
\begin{equation}
  \eta(\tau)\defeq \me^{\frac{\pi\ii}{12}\tau}\prod_{n=1}^{\infty}(1-\me^{2\pi\ii n \tau}).
  \label{}
\end{equation}
We have (a well-known calculation)
\begin{align}
  \sum_{n=1}^{\infty} \log \left \vert 1-\me^{2\pi\ii n\tau} \right \vert & \le \sum_{n=1}^{\infty}\sum_{m=1}^{\infty}\frac{1}{m}|\me^{2\pi\ii mn\tau}|=
  \sum_{n=1}^{\infty}\sum_{m=1}^{\infty}\frac{1}{m}\me^{-2\pi mn\fk{Im}(\tau)}\\
  &=\sum_{m=1}^{\infty}\frac{1}{m}\frac{\me^{-2\pi m\fk{Im}(\tau)}}{1-\me^{-2\pi m\fk{Im}(\tau)}} \le 
  C_{\alpha}\sum_{m=1}^{\infty}\me^{-2\pi m\fk{Im}(\tau)}\\
  &=\frac{C_{\alpha}\me^{-2\pi\fk{Im}(\tau)}}{1-\me^{-2\pi\fk{Im}(\tau)}}
  \label{eqn-estimate-eta-function}
\end{align}
whenever~$\fk{Im}(\tau)\ge \alpha>0$ for some~$\alpha$. In particular we have 
$$ \lim_{\fk{Im}(\tau)\rightarrow +\infty}\sum_{n=1}^{\infty} \log \left \vert 1-\me^{2\pi\ii n\tau} \right \vert  =0.$$
Note that we have 
$$\mathrm{Vol}( \mb{C}/\Lambda_{\tau})=\fk{Im}(\tau),$$
therefore we obtain directly
$$\lim_{\fk{Im}(\tau)\rightarrow +\infty} \frac{\log \detz(\Delta_{\mb{C}/\Lambda_{\tau}})}{\mathrm{Vol}( \mb{C}/\Lambda_{\tau})} 
=\lim_{\fk{Im}(\tau)\rightarrow +\infty} \frac{4\log \vert \eta(\tau)\vert}{\fk{Im}(\tau)}=\frac{-\pi}{3}.$$
A relevant special case is when $\tau=iLN$ with $L>0$, where $\mb{C}/\Lambda_{iNL}$ is an $N$-cyclic cover of $\mb{C}/\Lambda_{iL}$.
This explicit example shows that, in contrast with graphs, the limit of Theorem \ref{thm:convergencewithmass} is in general non-vanishing.

\subsubsection{Determinants of Laplacians on hyperbolic surfaces}
In what follows $X=\Gamma \backslash \mathbb{H}$ is a compact hyperbolic surface given by a quotient of the Poincar\'e half-plane $\mathbb{H}$ by a co-compact Fucshian group $\Gamma$. If $g$ denotes the genus of $X$, we know by Gauss-Bonnet that $\mathrm{Vol}(X)=4\pi(g-1)$. We denote by
$\Delta_X$ the hyperbolic Laplacian on $X$. In addition, let $\mathscr{M}_g$ denote the moduli space of compact hyperbolic surfaces with genus $g$. It follows
\footnote{Warm thanks to Peter Sarnak for pointing out the analogy with McKay's law for graphs, in a private communication.}
 from \cite{Na23} that we have the ``archimedian" McKay law:
$$\limsup_{g\rightarrow +\infty \atop X\in \mathscr{M}_g} \frac{\log \detz(\Delta_X)}{\mathrm{Vol}(X)}=E,$$
where we have set
$$E=\frac{4\zeta'(-1)-1/2+\log(2\pi)}{4\pi}\simeq 0.0538.$$
In addition, this limsup is actually a genuine limit for typical sequences of random surfaces and arithmetic congruence covers, see \cite{Na23} for precise statements. Needless to say, cyclic or abelian covers of Hyperbolic surfaces are highly non-typical sequences and in this setting, Theorem \ref{thm:convergencewithmass} is new.
It would be interesting to know if this limit is actually smaller than $E$ as in the analog case of graphs.

\paragraph{Acknowledgement.} We sincerely acknowledge Junrong Yan for suggesting the elegant estimate presented in the proof of Lemma \ref{lemm-rough-counting} which greatly improved an earlier method. We thank IMJ-PRG and Universit\'e de Strasbourg for supporting this research.

\section{Main Strategy and Proof}
The spectral zeta function of a positive semi-definite elliptic differential operator~$P$, with strictly positive principal symbol\footnote{This means the principal symbol~$\sigma_P(x,\xi)$ is positive definite for~$\xi\ne 0$ (\cite{Gilkey} section 1.6.2).}, acting on a Hermitian vector bundle over a Riemannian manifold (\cite{Gilkey} section 1.12) reads~:
\begin{equation}
  \zeta_P(s)=\frac{1}{\Gamma(s)}\int_{0}^{\infty}t^{s-1}\big(\ttr_{L^2}(\me^{-t P})-\dim\ker P \big)\,\dd t.
  \label{eqn-general-mellin-zeta-func}
\end{equation}
We denote by~$\me^{-tP}(x,y)$ the heat kernel which is smooth on~$M\times M$. In the first part of this paper, we consider~$P=\Delta_{\infty}$,~$\Delta_N$, or~$\Delta$, the Laplacians on~$M_{\infty}$,~$M_N$ or~$M$ respectively, acting on functions. In the last section~\ref{s:bundlecase}, we treat the case of Bochner Laplacians on certain flat Hermitian vector bundles. 

By definition of the~$\zeta$-determinant, to prove the existence of the limit in theorem \ref{thm:convergencewithmass}, it suffices to show that~$\{\zeta_{\Delta_N}(s)/\vol(M_N)\}$ is a sequence of holomorphic functions converging in the compact-open topology on a neighborhood of~$s=0$ as~$N\to +\infty$. To do this, 
we use the crucial fact that the heat kernels on~$M_N$ and~$M_{\infty}$ are related in a simple way by summing over the deck orbits~\cite[section 7.5]{Buser}. 
Applying a Li-Yau type estimate on the heat kernel over $M_{\infty}$, we obtain the convergence of~$N^{-1}\ttr_{L^2(M_N)}(\me^{-t\Delta_N})$ as~$N\to +\infty$ for fixed~$t>0$. 
The main technical part then boils down to obtaining a bound on the heat trace in~$t$, uniform in~$N$, to justify dominated convergence.

\subsection{Relation of Heat Kernels with Li-Yau Estimates}

The infinite cover $M_\infty$ endowed with the metric induced from $(M,g)$ is a geodesically complete Riemannian manifold by the Hopf-Rinow Theorem since every closed bounded subset in it is compact.
Denote by $\Delta_{\infty}$ the Laplacian on $M_\infty$, $\pi_N:M_\infty\lto M_N$. The geodesic completeness implies that $\Delta_{\infty}:C^\infty_c(M_\infty)\subset L^2(M_\infty)\lto L^2(M_\infty) $ has a self-adjoint extension by the work of Gaffney. 
Then observe we have the identity
\begin{eqnarray*}
\left(\pi_N\times \pi_N\right)^* e^{-t(\Delta_N+1)}=\sum_{k\in \mathbb{Z}} e^{-t(\Delta_\infty+1)}(.,\gamma^{kN}.).
\end{eqnarray*}
We make the crucial observation that
\begin{eqnarray*}
\ttr_{L^2(M_N)}(e^{-t\Delta_N})=\sum_{k\in \mathbb{Z}}\int_{\Omega_N}e^{-t\Delta_\infty}(x,\gamma^{kN}x)\,\dd V_g(x)\,=
N\sum_{k\in \mathbb{Z}}\int_{\Omega}e^{-t\Delta_\infty}(x,\gamma^{kN}x)\,\dd V_g(x)\,
\end{eqnarray*}
where $\Omega$, $\Omega_N$ are the respective fundamental domains of $M,M_N$ on the cover $M_\infty$, we also used the invariance equation $e^{-t\Delta_\infty}(x,x)=e^{-t\Delta_\infty}(\gamma(x),\gamma(x))$ which follows from the isometric action of $\gamma:M_\infty\mapsto M_\infty$.

In the sequel, we will make an extensive use of the following Lemma
which is a particular case of Milnor-Schwarz~\cite[Prop 8.19 p.~140]{BrHa}:
\begin{lemm}[Milnor-Schwarz type Lemma]
Let $(M,g)$ be a smooth compact Riemannian manifold and $M_\infty\lto M$ a $\mathbb{Z}$ cover of $M$. Then
there exists $L>0$ such that for all $x\in M_\infty$:
$$ \textbf{d}(x,\gamma^px)\geqslant pL. $$
A similar kind of estimate holds true in the case of $\mathbb{Z}^d$ covers.
\end{lemm}

The second ingredient we shall need are Gaussian bounds on the heat kernel $e^{-t(\Delta_\infty+1)}$ which are due to Li-Yau~\cite[Thm 4.6 p.~169]{SchoenYau}:
\begin{thm}[Li-Yau heat kernel bounds]
Let $(M_\infty,g)$ be a complete Riemannian manifold of dimension $d$ with $\mm{Ric}(M)\ge -K$, $K\ge 0$. Then for all $0<t\le 1$, $\delta\in (0,1)$ and $(x,y)\in M_\infty\times M_{\infty}$, the heat kernel $e^{-t\Delta_\infty}$
satisfies the bound of the form
\begin{eqnarray}
\vert e^{-t\Delta_\infty}(x,y)\vert\leqslant C_1(\delta,d)t^{-\frac{d}{2}} e^{-\textbf{d}^2(x,y)/(4+\delta)t}\me^{C_3 K\delta t}
\label{eqn-li-yau-bound}
\end{eqnarray}
where $\textbf{d}$ is the Riemannian distance and $C_1(\delta,d)\rightarrow +\infty$ when $\delta\rightarrow 0$, and $C_3$ depends only on the dimension.
\end{thm}

\subsection{Proof of Theorem \ref{thm:convergencewithmass}.}
Here we prove theorem \ref{thm:convergencewithmass} while leaving details on the estimate of the heat trace to section \ref{sec-long-time-control}.

\begin{proof}
    (Proof of theorem \ref{thm:convergencewithmass}.) Using (\ref{eqn-general-mellin-zeta-func}), we decompose
    \begin{equation}
        \frac{\zeta_{\Delta_N}(s)}{\vol(M_N)}=\frac{1}{\Gamma(s)\vol(M_N)}\Big( \underbrace{\int_0^1}_A +\underbrace{\int_1^{\infty}}_B\Big) \big(\ttr_{L^2(M_N)}(e^{-t\Delta_N})-1\big)t^{s-1}\,\dd t.
    \end{equation}
    For the part $A$, we have
     \begin{equation}
    \frac{1}{N}\int_{0}^{1}\big(\ttr_{L^2(M_N)}(e^{-t\Delta_N})-1\big)t^{s-1}\,\dd t =
    \int_{0}^{1}\Big[\int_{\Omega}^{} \sum_{k\in\mb{Z}} \me^{-t\Delta_{\infty}}(x,\gamma^{kN}x)\,\dd V_g(x) -\frac{1}{N}\Big] t^{s-1}\,\dd t.
    \label{}
  \end{equation}
  We observe by the Milnor-Schwarz lemma and Li-Yau estimate (and positivity of the heat kernel) that
    \begin{align*}
|\mathcal{T}_{\ne 0}(t,s)|&\defeq \bigg| t^{s-1} \int_{\Omega} 
\sum_{\substack{k\in \mathbb{Z}\\ k\neq 0}}e^{-t\Delta_\infty}(x,\gamma^{kN}x)\,\dd V_g(x)\bigg| \leqslant 
   \vert t^{s-1} \vert \,\int_{\Omega}
  \sum_{\substack{k\in \mathbb{Z}\\ k\neq 0}}e^{-t\Delta_\infty}(x,\gamma^{kN}x)\,\dd V_g(x)\\
  &\lesssim 
  t^{\fk{Re}(s)-1-\frac{d}{2}} \int_{\Omega}  \sum_{\substack{k\in \mathbb{Z}\\ k\neq 0}}e^{-C_2 \mathbf{d}^2(x,\gamma^{kN}(x))/t} \,\dd V_g(x) \\
& \lesssim  \vol(M) \cdot t^{\fk{Re}(s)-1-\frac{d}{2}}\Big(\sum_{\substack{k\in \mathbb{Z}\\ k\neq 0}} e^{-C_2 (kNL)^2 /t}\Big)    \le  \vol(M) \cdot t^{\fk{Re}(s)-1-\frac{d}{2}}   \Big(2\sum_{k=1}^{\infty} e^{-C_2 kNL /t}\Big)   \\
& = \vol(M) \frac{ 2e^{-C_2 NL/t}}{1-e^{-C_2 NL/t}}\cdot t^{\fk{Re}(s)-1-\frac{d}{2}}  
\lesssim e^{-C_2 NL/t} \cdot t^{\fk{Re}(s)-1-\frac{d}{2}}.\tag{\#}
\end{align*}
 Since $ e^{-\frac{C}{t}} t^{\fk{Re}(s)-1-\frac{d}{2}} =\mathcal{O}(t^\infty)$ for $C>0$ as $t\downarrow 0$, the r.h.s.\ of the above is obviously integrable on $(0,1)$, regardless of $s$. This shows firstly $\int_0^1 \mathcal{T}_{\ne 0}(t,s)\,\dd t$ in fact defines an entire function of $s$, and secondly from a simple change of variables we have $\int_0^1 e^{-C_2 NL/t} \cdot t^{\fk{Re}(s)-1-\frac{d}{2}}\,\dd t \lesssim N^{\fk{Re}(s)-1-\frac{d}{2}}$, and we conclude that
$$  \int_0^1 \mathcal{T}_{\ne 0}(t,s) \,\dd t
\lesssim N^{\fk{Re}(s)-1-\frac{d}{2}}\quad \xlongrightarrow{N\to +\infty} \quad 0$$
for $\fk{Re}(s)\ll 0$ and hence as entire functions of $s$ (Montel-Vitali theorem). Since
  \begin{equation}
    \frac{1}{N\Gamma(s)}\int_{0}^{1}t^{s-1}\,\dd t=\frac{1}{N\Gamma(s+1)}\quad \xlongrightarrow{N\to +\infty} \quad 0,
    \label{}
  \end{equation}
  we deduce
  \begin{equation}
    \frac{1}{N\Gamma(s)}\int_{0}^{1}\big(\ttr_{L^2(M_N)}(e^{-t\Delta_N})-1\big)t^{s-1}\,\dd t\quad \xlongrightarrow{N\to +\infty} \quad  
    \frac{1}{\Gamma(s)}\int_{0}^{1}t^{s-1}\int_{\Omega}^{} \me^{-t\Delta_{\infty}}(x,x) \,\dd V_g(x)\,\dd t,
    \label{}
  \end{equation}
  the latter defining a meromorphic function of~$s$ which is holomorphic near~$s=0$ by the asymptotic expansion of the heat kernel (\cite{BGV} theorem 2.30) and that~$\vol(\Omega)=\vol(M)<\infty$.

  To treat part~$B$, we first note that by the same estimates (\#) we have shown
\begin{equation}
  \lim_{N\to +\infty} \frac{1}{N}\big(\ttr_{L^2(M_N)}(e^{-t\Delta_N})-1\big) =\int_{\Omega}^{} \me^{-t\Delta_{\infty}}(x,x) \,\dd V_g(x)
  \label{eqn-heat-trace-pointwise-limit}
\end{equation}
for each fixed~$t>0$. Now by lemma \ref{lemm-long-time-bound-heat-trace} and dominated convergence we obtain
\begin{equation}
     \lim_{N\to +\infty}\frac{1}{N\Gamma(s)}\int_{1}^{\infty}\big(\ttr_{L^2(M_N)}(e^{-t\Delta_N})-1\big)t^{s-1}\,\dd t=  \frac{1}{\Gamma(s)}\int_{1}^{\infty}t^{s-1}\int_{\Omega}^{} \me^{-t\Delta_{\infty}}(x,x) \,\dd V_g(x)\,\dd t,
\end{equation}
as holomorphic functions of $s$ for $\fk{Re}(s)<1/2p$ and hence in particular near $s=0$. Combining the two parts, we obtain the result with limit
\begin{equation}
  \lim_{N\to +\infty}\frac{\zeta_{\Delta_N}(s)}{\vol(M_N)}=\frac{\zeta_{\Delta_{\infty},\Omega}(s)}{\vol(M)}
  \label{}
\end{equation}
where
\begin{equation}
  \zeta_{\Delta_{\infty},\Omega}(s)\defeq \frac{1}{\Gamma(s)}\int_{0}^{\infty}t^{s-1}\int_{\Omega}^{} \me^{-t\Delta_{\infty}}(x,x) \,\dd V_g(x)\,\dd t.
  \label{}
\end{equation}
Note that we know a posteriori after (\ref{eqn-heat-trace-pointwise-limit}) and lemma \ref{lemm-long-time-bound-heat-trace} that~$\zeta_{\Delta_{\infty},\Omega}(s)$ is a holomorphic function of~$s$ near~$s=0$ as the sum of two such functions, one being a meromorphic continuation.
\end{proof}

\subsection{Accumulation of Eigenvalues and Long-time Behavior of Heat Trace}\label{sec-long-time-control}

The goal of this subsection is to prove
\begin{lemm}\label{lemm-long-time-bound-heat-trace}
  There exists a positive integer~$p$, a number~$\varepsilon_0>0$, and constants~$C_4$,~$C_5>0$ independent of~$N$, such that 
  \begin{equation}
    \frac{1}{N}\big(\ttr_{L^2(M_N)}(e^{-t\Delta_N})-1\big) \le C_4 t^{-1/2p} +C_5 \me^{-t \varepsilon_0},
    \label{}
  \end{equation}
  for all $N\in\mb{N}$ and all $t\ge 1$.
\end{lemm}

\begin{proof}
  We obtain from corollary \ref{cor-small-eigen-estima} the threshold~$\varepsilon_0>0$,~$p$ and~$C_4$ independent of~$N$ so that
  \begin{equation}
    \frac{1}{N}\sum_{\substack{\lambda\in \sigma(\Delta_{N}) \\ 0<\lambda<\varepsilon_0}} \me^{-t\lambda} \le C_4 t^{-1/2p}.
    \label{}
  \end{equation}
  Now since~$\vol(M_N)=N\vol (M)$, and~$\mm{Ric}(M_N)\ge \mm{Ric}(M)$, by lemma \ref{lemm-rough-counting} we have some~$C_6$ independent of~$N$ such that
  \begin{equation}
  \sharp\{\lambda\le \Lambda|\lambda\in \sigma(\Delta_N)\}\le C_6 N\Lambda^{d/2}.
  \label{}
\end{equation}
Now pick~$\varepsilon_0\ll\Lambda_0$, and put $\Lambda_j:=\Lambda_0+j$. Then we have for~$t\ge 1$,
\begin{align*}
  \sum_{\substack{\lambda\in \sigma(\Delta_N) \\ \lambda\ge \varepsilon_0}} \me^{-t\lambda}&\le \sharp\big([\varepsilon_0,\Lambda_0)\cap \sigma(\Delta_N)\big) \,\me^{-t\varepsilon_0}+\sum_{j=0}^{\infty}\sharp\big([\Lambda_j,\Lambda_{j+1})\cap \sigma(\Delta_N)\big)\, \me^{-t\Lambda_j}\\
  &\le C_6 N\Lambda_0^{d/2}\me^{-t\varepsilon_0}+\me^{-t\Lambda_0}\sum_{j=0}^{\infty} C_6 N\Lambda_{j+1}^{d/2} \me^{-tj}\\
  &\le C_6 N\Lambda_0^{d/2}\me^{-t\varepsilon_0}+ C_6 N \me^{-t\Lambda_0} \underbrace{\sum_{j=0}^{\infty} (\textrm{poly. in }j)\cdot  \me^{-j}}_{\textrm{converge}} \\
  &\le C_5 N\me^{-t\varepsilon_0}.
\end{align*}
We obtain the result.
\end{proof}

\subsection{Twisted Laplacians and Decomposition of Spectra}

The next step is to identify the action of the Laplace operator $\Delta_N$ on $M_N$ with the action of some family of Laplacians acting on functions on the base manifold $M$.

\subsubsection{Fourier Transform in the Fibers}

Consider $M_N\lto M
$ the $N$-th degree cyclic cover, we may think of the cover $M_N\lto M$ as a $\mathbb{Z}/N\mathbb{Z}$--principal bundle over $M$ so that we can use the discrete Fourier transform over the fibers. Then we
observe that
given any function $f$ on $M_N$, the deck group of $M_N$ acts on the function $f$ as some discrete rotation of the fibers. Therefore,
we can define the \textsf{fiberwise Fourier transform} $\wh{\bullet}:C^{\infty}(M_N)\lto C^{\infty}(M_N)^N$ by
\begin{eqnarray*}
\widehat{f}_p(x)\defeq \sum_{k=1}^N f(\gamma_0^k.x)e^{-2\pi \ii\frac{ p}{N}k},\quad\quad\textrm{for }0\le p\le N-1.
\end{eqnarray*}
where $\gamma_0$ is the \textit{generator} of the deck shifts. The Fourier inversion formula subsequently reads
\begin{eqnarray}
\label{eq:automorphycondition}
f(x)=\frac{1}{N}\sum_{p=1}^N \widehat{f}_p(x), \quad\quad x\in M_N.
\label{eqn-automorphic-fiber-fourier}
\end{eqnarray}
Now we define several function spaces as follows.
\begin{align}
  L^2_{p,N}(M_N)&\defeq \{u\in L^2(M_N)~|~ \gamma_0^* u=\me^{2\pi\ii \frac{p}{N}}u\},\\
  C^{\infty}_{p,N}(M_N)&\defeq \{u\in C^{\infty}(M_N)~|~ \gamma_0^* u=\me^{2\pi\ii \frac{p}{N}}u\},\\
  C^{\infty}_{p,N}(\Omega)&\defeq \{u|_{\Omega}~|~ u\in C^{\infty}_{p,N}(M_N)\},
  \label{}
\end{align}
where~$\Omega$ is the fundamental domain of $M$ in $M_\infty$. Observe that $\wh{f}_p\in L_{p,N}^2(M_N)$ and
\begin{lemm}
  We have~$L^2_{p,N}(M_N)\perp L^2_{q,N}(M_N)$ in~$L^2(M_N)$ for~$p\ne q$.
\end{lemm}

\begin{proof}
    We have for $u\in L^2_{p,N}$ and $v\in L^2_{q,N}$, since the deck shifts are transitive,
    \begin{align*}
\left\langle u,v\right\rangle_{L^2(M_N)}&= \sum_{k=1}^N  \int_{\Omega} \overline{u}(\gamma^k.x)v(\gamma^k.x)\,\dd V_g(x)\,\\
&=
\sum_{k=1}^N e^{-2\pi \ii\frac{ p}{N}k}e^{2\pi \ii\frac{ q}{N}k} \int_{\Omega} \overline{u}(
x)v( x)\,\dd V_g(x)=0,
\end{align*}
by the automorphy condition and since $\sum_{k=1}^N e^{2\pi \ii\frac{ p-q}{N}k}=0 $.
\end{proof}

The Fourier inversion formula thus gives the orthogonal decomposition
\begin{equation}
  L^2(M_N)=\bigoplus_{p=0}^{N-1}L_{p,N}^2(M_N).
  \label{}
\end{equation}
Note that since~$\gamma_0$ acts by isometry we have~$\Delta_N:C^{\infty}_{p,N}(M_N)\lto C^{\infty}_{p,N}(M_N)$ for each~$p$. Since~$\Delta_N$ is essentially self-adjoint on~$C^{\infty}(M_N)$, we have also decomposed
\begin{equation}
  \Delta_N=\bigoplus_{p=0}^{N-1} \Delta_N|_{L_{p,N}^2(M_N)}.
  \label{}
\end{equation}

Next we observe that by the automorphy condition, the action of~$\Delta_N$ on each~$L_{p,N}^2(M_N)$ boils down to its action on~$L^2(\Omega)$ with a specified boundary condition. More precisely, given~$f\in L^2(\Omega)$, define its \textsf{(quasi-)periodization}~$\mathcal{P}_{p,N}f\in L^2(M_N)$, such that as a distribution
\begin{equation}
  \bank{\mathcal{P}_{p,N}f, \varphi}_{L^2(M_N)}\defeq \bank{f,\wh{\varphi}_p|_{\Omega}}_{L^2(\Omega)},
  \label{}
\end{equation}
for test functions~$\varphi\in C^{\infty}(M_N)$. One can subsequently verify that indeed~$\mathcal{P}_{p,N}f\in L^2_{p,N}$ by resorting to the adjoint of~$\gamma_0^*$. Moreover,~$\mathcal{P}_{p,N}:C^{\infty}_{p,N}(\Omega)\lto C^{\infty}_{p,N}(M_N)$. We arrive at the main observation of this subsection.
\begin{lemm}\label{lemm-unit-equiv-domain-bdy-cond}
  For each integer~$0\le p\le N-1$ we have the commutative diagram
  \begin{equation}
    \begin{tikzcd}
      C^{\infty}_{p,N}(\Omega) \ar[r," \Delta_{\Omega}"] \ar[d,"\mathcal{P}_{p,N}"'] & L^2(\Omega) \ar[d,"\mathcal{P}_{p,N}"] \\[+10pt]
      C^{\infty}_{p,N}(M_N) \ar[r," \Delta_N"] & L^2_{p,N}(M_N).
    \end{tikzcd}
    \label{}
  \end{equation}
  Denote by~$\Delta_{p,N}$ the self-adjoint extension of the Laplacian on~$L^2(\Omega)$ with core~$C^{\infty}_{p,N}(\Omega)$. Then~$\Delta_{p,N}$ is unitarily equivalent to~$\Delta_N|_{L_{p,N}^2(M_N)}$ through~$\frac{1}{N}\mathcal{P}_{p,N}$ for each~$0\le p\le N-1$. 
\end{lemm}

\begin{proof}
  We only point out that in this case the existence and uniqueness of the self-adjoint extension~$\Delta_{p,N}$ can be obtained thanks to the unitary map~$\frac{1}{N}\mathcal{P}_{p,N}$ and the corresponding result for~$\Delta_N|_{L_{p,N}^2(M_N)}$. A more general dituation is considered in lemma \ref{lemm-twist-self-adjoint-domain}.
\end{proof}

\subsubsection{Characters and Unitary Equivalence}

Our cover~$M_{\infty}$ was constructed to correspond to the normal subgroup of~$\pi_1(M)$ which is the kernel of the map
  \begin{equation}
  \begin{tikzcd}
    \rho:\pi_1(M) \ar[r,"\mm{Ab}"] &[+10pt] H_1(M;\mb{Z}) \ar[r, " I(-{,}{[}\Sigma{]})"] &[+10pt] \mb{Z},
  \end{tikzcd}
  \label{}
\end{equation}
where the first is Abelianization and~$I(-,[\Sigma])$ is the \textsf{oriented intersection number} with some cycle~$[\Sigma]\in  H_{n-1}(M,\mathbb{Z})$, the surjectivity of $\rho$ implies the class $[\Sigma]\in H_{n-1}(M,\mathbb{Z})$ is non torsion. 
By Poincar\'e duality and Hodge theory on~$M$, there exists a unique harmonic $1$-form~$\alpha_{\Sigma}$ such that
\begin{equation}
  I([\gamma],[\Sigma])=\int_{\gamma}^{}\alpha_{\Sigma}
  \label{eqn-def-1-form-inter-number}
\end{equation}
for all (smooth) loops~$\gamma$ in~$M$. By the construction of the cover~$\pi_{\infty}:M_{\infty}\lto M$, there is the relation
\begin{equation}
  (\pi_{\infty})_*(\pi_1(M_{\infty},x_0))=\ker \rho,
  \label{}
\end{equation}
upon picking a base point~$x_0\in M_{\infty}$ and interpreting~$\pi_1(M)$ as~$\pi_1(M,\pi_{\infty}(x_0))$. This translates into saying that
\begin{equation}
  \int_{(\pi_{\infty})_*\tilde{\gamma}}^{}\alpha_{\Sigma}=\int_{\tilde{\gamma}}^{} \pi_{\infty}^* \alpha_{\Sigma}=0
  \label{}
\end{equation}
for all loops~$\tilde{\gamma}$ in~$M_{\infty}$ based at~$x_0$. This means~$\pi_{\infty}^* \alpha_{\Sigma}\in \Omega^1(M_{\infty})$ is exact ($M_{\infty}$ is connected), and a primitive is given by
\begin{equation}
  \psi_{\Sigma}(x)\defeq \int_{x_0}^{x} \pi_{\infty}^* \alpha_{\Sigma}.
  \label{eqn-def-harmonic-primitive}
\end{equation}
Observe moreover that~$\psi_{\Sigma}$ is harmonic since~$\alpha_{\Sigma}$ is harmonic (thus co-closed). The most important property of~$\psi_{\Sigma}$ for us is the following which we single out as a lemma.

\begin{lemm}
  For~$\gamma\in \pi_1(M)$, use the same notation to denote the deck action of the class $\ol{\gamma}\in \pi_1(M)/\ker\rho$ on~$M_{\infty}$. Then we have
  \begin{equation}
    \psi_{\Sigma}(\gamma.x)=I([\gamma],[\Sigma])+\psi_{\Sigma}(x)=\rho(\gamma)+\psi_{\Sigma}(x),
    \label{eqn-deck-auto-primitive-potential}
  \end{equation}
  for any~$x\in M_{\infty}$.
\end{lemm}

\begin{proof}
  This comes from lifting loops, (\ref{eqn-def-harmonic-primitive}) and the property (\ref{eqn-def-1-form-inter-number}).
\end{proof}

Now we introduce several helpful functions. Let~$\theta\in [-\pi,\pi)$ (identified with the torus) be a parameter. Define the \textsf{character}
\begin{equation}
  \left.
  \begin{array}{rcl}
    \chi_{\theta}: \pi_1(M)&\lto & \mb{C},\\
    \gamma &\longmapsto & \me^{\ii \theta \int_{\gamma}^{}\alpha_{\Sigma}}= \me^{\ii \theta\cdot  \rho(\gamma)}.
  \end{array}
  \right.
  \label{}
\end{equation}
Next we define for $\theta\in [-\pi,\pi)$ the simple automorphic function~$G_{\theta}:M_{\infty}\lto \mb{C}$,
\begin{equation}
  G_{\theta}(x)\defeq\me^{\ii \theta\psi_{\Sigma}}.
  \label{}
\end{equation}
By (\ref{eqn-deck-auto-primitive-potential}) we find that
\begin{equation}
  G_{\theta}(\gamma.x)=\chi_{\theta}(\gamma)G_{\theta}(x),
  \label{}
\end{equation}
for deck shifts~$\gamma\in \pi_1(M)$,~$x\in M_{\infty}$. Now we restrict~$G_{\theta}$ to the fundamental domain~$\Omega$. Similarly as before we define
\begin{equation}
  C_{\theta}^{\infty}(\Omega)\defeq \left\{\tilde{f}|_{\Omega}~ \middle|~ \tilde{f}\in C^{\infty}(M_{\infty}),~\tilde{f}(\gamma.x)=\chi_{\theta}(\gamma)\tilde{f}(x)\textrm{ for }\gamma\in \pi_1(M) \right\}.
  \label{}
\end{equation}
Denote by~$\Delta_{\Omega,\theta}$ the self-adjoint Laplacian acting on~$L^2(\Omega)$ with core~$C_{\theta}^{\infty}(\Omega)$, sometimes called a \textsf{Born-von K\'arm\'an Laplacian} \cite{Lewin}. Note that~$C_0^{\infty}(\Omega)$ is identified with~$C^{\infty}(M)$, and trivially~$L^2(\Omega)$ with~$L^2(M)$. 

\begin{lemm}\label{lemm-twist-self-adjoint-domain}
  The Laplacian is indeed essentially self-adjoint on~$C_{\theta}^{\infty}(\Omega)$. The map~$G_{\theta}\mn{\cdot}:L^2(\Omega)\lto L^2(\Omega)$,~$f\mapsto G_{\theta}f$ is a unitary isomorphism. The operator~$\Delta_{\theta}$ on~$L^2(M)$ defined by~$\Delta_{\theta}f:=G_{\theta}^{-1}\Delta_{\Omega,\theta}(G_{\theta}f)$ is essentially self-adjoint on the core~$C^{\infty}(M)$ and unitarily equivalent to~$\Delta_{\Omega,\theta}$.
\end{lemm}

Note that if we identify~$M\setminus \Sigma$ with~$\Omega^{\circ}$, then~$G_{\theta}$ has monodromy crossing the cut~$\Sigma$. However, for~$f\in C^{\infty}(M)$ it follows from local computations with~$\Delta=-\ddiv \nabla$ that
\begin{equation}
  \Delta_{\theta}f=\Delta_M f -2\ii\theta \alpha_{\Sigma}( \nabla f)+\theta^2|\alpha_{\Sigma}|_g^2 f,
  \label{eqn-express-twist-lap}
\end{equation}
where~$\alpha_{\Sigma}=\dd \psi_{\Sigma}$ is the harmonic 1-form on~$M$ obtained in (\ref{eqn-def-1-form-inter-number}). Thus we do have~$\Delta_{\theta}f\in C^{\infty}(M)$.\\

\begin{proof}(Proof of lemma \ref{lemm-twist-self-adjoint-domain}.)
  We point out that~$\Delta_{\Omega,\theta}$ will have domain
  \begin{equation}
    \dom(\Delta_{\Omega,\theta})=\left\{ 
      f\in H^2(\Omega)~\middle| 
      \left.
      \begin{array}{l}
	f|_{\Sigma_+}=\me^{\ii\theta}f|_{\Sigma_-},\\
	(\partial_{\nu_+}f)|_{\Sigma_+}=\me^{\ii\theta}(\partial_{\nu_-}f)|_{\Sigma_-}
      \end{array}
      \right.
    \right\},
    \label{eqn-domain-boundary-cond-von-karman}
  \end{equation}
  where~$H^2(\Omega)$ denotes the second order Sobolev space (\cite{Lewin} page 85 for the torus case). We also see from (\ref{eqn-express-twist-lap}) that~$\dom(\Delta_{\theta})=H^2(M)$. Consequently it's clear that~$G_{\theta}\cdot \dom(\Delta_{\theta})=\dom(\Delta_{\Omega,\theta})$. The rest of the lemma follows.
\end{proof}

Combining lemmas \ref{lemm-unit-equiv-domain-bdy-cond} and \ref{lemm-twist-self-adjoint-domain}, we obtain
\begin{equation}
  \sigma(\Delta_N)=\bigcup_{p=0}^{N-1}\sigma(\Delta_{2\pi p/N})=
  \bigcup_{p=0}^{N-1}\sigma(\Delta_{p,N}).
  \label{eqn-main-spectral-decomp}
\end{equation}

Such operators already appeared in the works~\cite{PS,KS} whose purpose was to count geodesics with homological constraints, as well as in the more recent works~\cite[Appendix]{AnETDS}, \cite[Remark 1.10 p.~600]{AnGAFA} where $\Delta_{p,N}$ is called \emph{twisted Laplacians}~\cite[p.~599, 610]{AnGAFA}.

\subsubsection{Vanishing of Bottom of Spectrum}

%
%
%
%

\begin{lemm}\label{lemm:minimaslambda0}
  Let~$\Delta_{\theta}$ be the operator defined in lemma \ref{lemm-twist-self-adjoint-domain} with~$\theta\in\mb{R}$. Then $\ker(\Delta_\theta)\neq \{0\}$ iff $\theta\in 2\pi\mathbb{Z}$.
\end{lemm}

\begin{proof}
  If~$\theta=2\pi k$, then~$G_{\theta}^{-1}=\me^{-2\pi\ii k\psi_{\Sigma}}\in C^{\infty}(M)\subset \dom(\Delta_{\theta})$. One could verify directly with (\ref{eqn-express-twist-lap}) that in this case~$G_{\theta}^{-1}\in\ker(\Delta_{\theta})$. Conversely, suppose now~$\ker(\Delta_{\theta})\ne \{0\}$. Here it turns out more convenient to work with the original operators~$\Delta_{\Omega,\theta}$ on the fundamental domain~$\Omega$. Indeed, for general~$s\in H^2(\Omega)$ we have by Green-Stokes formula ($\ol{s}$ denoting the complex conjugate of $s$)
  \begin{equation}
    \int_{\Omega}^{}\ol{s}\Delta_{\Omega}s\,\dd V_g=\int_{\Omega}^{}\sank{\ol{\nabla s},\nabla s}_g\,\dd V_g -\int_{\Sigma_+}^{} \ol{s}(\partial_{\nu_+}s)\,\dd V_g -\int_{\Sigma_-}^{}\ol{s}(-\partial_{\nu_-} s)\,\dd V_g,
    \label{eqn-green-stokes-funda-domain}
  \end{equation}
  where we remember that~$\nu_+$ is \textit{outward} pointing whereas~$\nu_{-}$ is \textit{inward} pointing. In particular, for~$s$ satisfying the boundary conditions in (\ref{eqn-domain-boundary-cond-von-karman}), the boundary terms of (\ref{eqn-green-stokes-funda-domain}) vanish. This shows that, under \textit{these boundary conditions},~$s\in \ker(\Delta_{\Omega,\theta})$ iff~$\nabla s=0$, which implies~$s$ is a constant. However, looking at the boundary conditions again, this could only happen with $s\ne 0$ if~$\theta\in 2\pi\mb{Z}$, proving the result.
\end{proof}

From (\ref{eqn-green-stokes-funda-domain}) and the boundary conditions one can also see that~$\Delta_{\Omega,\theta}$ are all nonnegative. On the other hand, from the expression (\ref{eqn-express-twist-lap}) for~$\Delta_{\theta}$ we see that it is a second order elliptic differential operator which by the usual methods has compact resolvent and hence discrete spectrum. Thus we denote the spectrum of~$\Delta_{\theta}$ by~$0\le \lambda_0(\theta)\le \lambda_1(\theta)\le \cdots$, counting multiplicity. Lemma \ref{lemm:minimaslambda0} then translates into saying that~$\lambda_0(\theta)=0$ iff~$\theta\in 2\pi\mb{Z}$, and~$\lambda_0(\theta)>0$ otherwise.

\subsection{Kato Perturbations and Uniform Spectral Gap}

Our goal is to control the bottom of the spectrum of the sequence $\Delta_N$ uniformly in $N$ when $N$ gets large enough. We have followed the approach of Phillips-Sarnak~\cite{PS} and obtained a family of operators $\Delta_{\theta}$ indexed by the parameter $\theta\in [-\pi,\pi)$ on the ``Jacobian torus'', which are related to the spectrum of $\Delta_N$ by (\ref{eqn-main-spectral-decomp}).

We have also shown that lowest eigenvalue $\lambda_0(\theta)$ of $\Delta_{\theta}$ vanishes exactly when $\theta=0$ for $\theta\in [-\pi,\pi)$. In this subsection we will apply Kato's perturbation theory to see that $\lambda_0(\theta)$ in fact depends analytically on $\theta$ near~$\theta=0$ and $\lambda_0(0)=0$ is a \textbf{strict minimum}.

\subsubsection{The Family of Twisted Laplacians Parametrized by the Angle}

\begin{defi}
    An (unbounded) operator-valued function~$T(\beta)$ on a complex (resp.\ real) domain~$\mathcal{R}$ is called an \textsf{analytic family (in the sense of Kato)} if
\begin{enumerate}[(i)]
  \item for every~$\beta\in \mathcal{R}$,~$T(\beta)$ is closed and has non-empty resolvent set~$\mb{C}\setminus \sigma(T(\beta))$;
  \item for every~$\beta_0\in \mathcal{R}$ there exists~$z_0\in \mb{C}\setminus \sigma(T(\beta_0))$ and~$\delta>0$ such that~$z_0$ remains in the resolvent of~$T(\beta)$ for~$|\beta-\beta_0|<\delta$, and~$(T(\beta)-z_0)^{-1}$ is a (resp.\ real) analytic operator-valued function of~$\beta$ for~$|\beta-\beta_0|<\delta$.
\end{enumerate}
\end{defi}

\begin{lemm}
  The function~$\theta\mapsto \Delta_{\theta}$ is a real analytic family on~$\mb{R}$.
\end{lemm}

For every $s\in \mathbb{R}$, we will denote by $\Psi^s(M)$ the pseudodifferential operators of order $s$ acting on distributions on the manifold $M$.\\

\begin{proof}
    Essentially, we see from the expression (\ref{eqn-express-twist-lap}) that~$H^2(M)$ is a \textit{common domain} for all of~$\Delta_{\theta}$,~$\theta\in \mb{R}$. We could verify with that expression also that for~$u\in H^2(M)$,~$\Delta_{\theta}u$ is a~$L^2(M)$-valued (real) analytic function of~$\theta$. These ensure that~$(\Delta_{\theta})_{\theta}$ is an \textit{analytic family of type (A)} in the sense of \cite{RS4} page 16 and is consequently a Kato-analytic family (see \cite{RS4} and Kato \cite{Kato} pages 375-381). 
    
    Nevertheless, we indicate the argument for our case to make the paper more self-contained. We would like to prove, say, the real analyticity near a fixed $\theta_0\in\mb{R}$. Clearly for any $L>0$, $-L\notin \sigma(\Delta_\theta)$ for all $\theta$ since $\Delta_\theta:H^2(M)\lto L^2(M)$ are self-adjoint with nonnegative spectrum. By definition of the operators $\Delta_\theta$, we have an identity of the form
\begin{eqnarray*}
\Delta_\theta=\Delta_M-\ii \theta X+\theta^2 V\defeq \Delta_M+A_\theta
\end{eqnarray*}
where $X$ is a smooth vector field (differential operator of order $1$) and $V$ a smooth potential on $M$. We would have
\begin{align*}
(\Delta_\theta+L)^{-1}=(\one+(L+\Delta)^{-1}A_\theta)^{-1}(L+\Delta)^{-1}
\end{align*} 
provided $\Vert
(L+\Delta)^{-1}A_\theta \Vert_{L^2\to L^2}<1$. Indeed, by the composition theorem for pseudodifferential operators, $(L+\Delta)^{-1}\circ A_\theta$ is bounded in $\Psi^{-1}(M)$ uniformly in $\theta$ near $\theta_0$ and therefore as operators in $\mathcal{L}(L^2(M))$ since $\Psi^{-1}(M)\subset \Psi^0(M)\subset \mathcal{L}(L^2(M))$ by the Calder\'on-Vaillancourt theorem. Thus by choosing $L$ large enough (possibly depending on $\theta_0$), we can make sure that
$ \Vert
(L+\Delta)^{-1}A_\theta \Vert_{L^2\to L^2}\leqslant \frac{1}{2}$, uniformly in $\theta$ near $\theta_0$ and the result follows by explicit power series expansion of $(\Delta_{\theta}+L(\theta_0))^{-1}$ in $\theta$ near $\theta_0$.
\end{proof}

\subsubsection{Analytic Vanishing of Bottom of Spectrum}

We have the next result which follows from application of the Kato--Rellich Theorem (see \cite{RS4} theorems XII.8 and XII.13):
\begin{lemm}\label{lemm:Kato}
Let $\Delta_\theta$ be the analytic family of operators defined in lemma \ref{lemm-twist-self-adjoint-domain}. There exists a threshold~$\varepsilon_0>0$ and an integer~$p>0$ such that
\begin{enumerate}[(i)]
  \item $[0,\varepsilon_0)$ contains no~$\lambda_i(\theta)$ for any~$i\ge 1$ and~$\theta\in[-\pi,\pi]$;
  \item there exists $a>b>0$ as well as~$\eta>0$ such that~$|\lambda_0(\theta)|=|\lambda_0(\theta)-\lambda_0(0)|<\varepsilon_0$ implies
  \begin{equation}
      b\theta^{2p}\le \lambda_0(\theta)=a\theta^{2p}+\mathcal{O}(\theta^{2p+1}), 
      \label{eqn-ground-energy-anal-local}
  \end{equation}
  as well as $|\theta|<\eta$, with $\lambda_0(0)=0$ being a strict minimum of $\lambda_0(\theta)$.
\end{enumerate}
\end{lemm}

\begin{proof}
  First of all we remark that~$\lambda_0(\theta)=\inf \sigma(\Delta_{\theta})$ is continuous in~$\theta$ since at every~$\theta_0$ we can choose~$L>0$ large enough so that~$-L$ is in the resolvent set of~$\Delta_{\theta}$ for all~$\theta$ near~$\theta_0$ and, as each $(\Delta_{\theta}+L)^{-1}$ is compact,~$(\lambda_0(\theta)+L)^{-1}=\sup \sigma((\Delta_{\theta}+L)^{-1})=\nrm{(\Delta_{\theta}+L)^{-1}}_{L^2\to L^2}$, the latter being continuous in~$\theta$ since $\Delta_{\theta}$ is a Kato-analytic family. This tells us that in fact
  \begin{equation}
    \inf_{\theta\in [-\pi,\pi]\setminus (-\delta,\delta)}\lambda_0(\theta)=\varepsilon_1>0
    \label{}
  \end{equation}
  for any~$0<\delta<\pi$ since otherwise it would contradict lemma \ref{lemm:minimaslambda0}. 
  Since we know that $\lambda_0(0)=0$ is a simple eigenvalue, each~$\Delta_{\theta}$ is nonnegative and we have lemma \ref{lemm:minimaslambda0}, we deduce by the Kato-Rellich theorem~\cite[Thms XII.8 and XII.13]{RS4} 
  that in some neighborhood~$|\theta|<\eta_1$,~$\theta=0$ is a strict minimum of the function~$\theta\mapsto \lambda_0(\theta)$, which is analytic in the neighborhood.
  Therefore in this neighborhood $\lambda_0(\theta)$ will have the local form on the r.h.s.\ of (\ref{eqn-ground-energy-anal-local}). Consequently for any $b<a$ the inequality on the l.h.s.\ of (\ref{eqn-ground-energy-anal-local}) will be satisfied when $|\theta|<\eta_3$ for some $\eta_3<\eta_1$.
  Next, since we also know that~$\lambda_1(0)>0$ is an isolated eigenvalue of finite multiplicity, by the finite multiplicity version of Kato-Rellich~\cite[Thm XII.13]{RS4}, we know that for some other~$\eta_2>0$,~$\lambda_1(\theta)$ is continuous in~$\theta$ for~$|\theta|<\eta_2$. Summing up,
   by necessarily choosing some
   \begin{equation}
       0<\varepsilon_0<\min\Big\{ \varepsilon_1, \inf_{|\theta|\le \eta_2}\lambda_1(\theta)\Big\},\quad \textrm{and}\quad
       \eta=\min\{\eta_2,\eta_3\},
   \end{equation}
   we will have (i) and (ii) satisfied for some~$p>0$ and we obtain the result.
\end{proof}

The spectral decomposition (\ref{eqn-main-spectral-decomp}) together with lemma \ref{lemm:Kato} tells us that for all~$N\in\mb{N}$, every~$\lambda\in [0,\varepsilon_0)\cap \sigma(\Delta_N)$ must be~$\lambda_0(2\pi k/N)$ for some integer~$k<\frac{1}{2}N$. This will give us the following important corollary.

\begin{coro}\label{cor-small-eigen-estima}
      Let~$\varepsilon_0$ and~$p$ be the numbers obtained in lemma \ref{lemm:Kato}. Then there exists~$C_4>0$ independent of~$N$, such that
  \begin{equation}
    \frac{1}{N}\sum_{\substack{\lambda\in \sigma(\Delta_{N}) \\ 0<\lambda<\varepsilon_0}} \me^{-t\lambda} \le C_4 t^{-1/2p}
    \label{}
  \end{equation}
  for all~$N\in\mb{N}$ and all $t>0$.
\end{coro}

 We point out the following formula related to the Gamma function:
  \begin{lemm}
    Let~$\alpha$,~$\beta>0$ then we have
    \begin{equation}
      \int_{0}^{\infty}\me^{-\beta x^{\alpha}}\,\dd x=\frac{1}{\alpha}\Gamma\Big( \frac{1}{\alpha}\Big) \beta^{-\frac{1}{\alpha}}.
      \label{}
    \end{equation}
  \end{lemm}

  \begin{proof}
      (Proof of corollary \ref{cor-small-eigen-estima}.) 
      By (ii) of lemma \ref{lemm:Kato} and the above formula, we have 
\begin{align*}
\frac{1}{N} \sum_{\substack{\lambda\in \sigma(\Delta_{N}) \\ 0<\lambda<\varepsilon_0}}  e^{-t \lambda}&\leqslant \frac{2\pi}{N} \sum_{0<2\pi|k|\leqslant \eta N}  e^{-t b(2\pi k/N)^{2p}}
\leqslant 2\int_{0}^\eta e^{-tb\theta^{2p}}\,\dd\theta \le \frac{1}{p}\Gamma\Big( \frac{1}{2p}\Big) (tb)^{-\frac{1}{2p}},
\end{align*}
the second inequality holds true since $e^{-u}$ is decreasing. We obtain the result.
  \end{proof}

\subsection{Rough Eigenvalue Counting}
\begin{lemm}\label{lemm-rough-counting}
  Let~$\mathcal{M}$ be a smooth compact Riemannian manifold of dimension~$d$ with~$\mm{Ric}(\mathcal{M})\ge -K$,~$K\ge 0$. Let~$\Delta_{\mathcal{M}}$ be the Laplacian on~$\mathcal{M}$. Then for all~$\Lambda\ge 1$ we have
  \begin{equation}
    \sharp\{\lambda\le \Lambda|\lambda\in \sigma(\Delta_{\mathcal{M}})\} \le 
    C_6\me^{1+C_3 K/2}\vol(\mathcal{M}) \Lambda^{d/2},
    \label{}
  \end{equation}
  where~$C_6$ is a constant depending only on dimension, and~$C_3$ the same constant appearing in (\ref{eqn-li-yau-bound}).
\end{lemm}

\begin{proof}
  Indeed, we have,
  \begin{equation}
    \sharp\{\lambda\le \Lambda|\lambda\in \sigma(\Delta_{\mathcal{M}})\}  =\sum_{\substack{\lambda\in \sigma(\Delta_{\mathcal{M}}) \\ 0\le \lambda\le \Lambda}} 1 \le \sum_{\substack{\lambda\in \sigma(\Delta_{\mathcal{M}}) \\ 0\le \lambda\le \Lambda}} \me^{1-\frac{1}{\Lambda}\lambda}
    \le \me \cdot \ttr_{L^2(\mathcal{M})}(\me^{-\Delta_{\mathcal{M}}/\Lambda}).
  \end{equation}
  Thus the result follows by integrating the Li-Yau estimate (\ref{eqn-li-yau-bound}) choosing $\delta=1/2$.
\end{proof}

\subsection{The Flat Vector Bundle Case}
\label{s:bundlecase}

We would like to generalize our previous result in the twisted case where we consider certain flat Hermitian vector bundle $p:(E,\nabla)\lto M$.
Recall from the introduction that we constructed our abelian covers by considering a certain group homomorphism $\rho:\pi_1(M)\mapsto\mathbb{Z}$. Now given any unitary matrix $U\in U(n)$, 
consider the morphism $ a\in \pi_1(M) \mapsto U^{\rho(a)} $ which yields a representation of the fundamental group  
$\pi_1(M)$ in $U(n)$. To such representation, we may functorially associate a flat bundle $E$ on $M$~\cite[Thm 13.2 p.~159]{Taubes}.
Then for every $N$, we can pull--back the bundle $E$ over $M_N$, we denote
by $E_N:=\pi_N^*E \lto M_N$ the corresponding pull-back bundle defined
as~:
$\pi_N^*E =\{ (x,v) \in M_N\times E | \pi_N(x)=p(v)    \}$.
From the definition, it is immediate that there exists a canonical action of the deck group $\mathbb{Z}_N$ on $E_N$ which canonically identifies fibers over $\mathbb{Z}_N$ orbits~:  
  $ (x,v),(\gamma.x,v),\dots,(\gamma^{N-1}.x,v)  $. 
Once the fibers  over every orbit are identified, we can write any section $s\in C^\infty(E_N)$ using the discrete Fourier transform as we did previously and decompose every section as we did before as a sum over Fourier modes. We decompose the spectrum of the Bochner Laplacian $\nabla^*\nabla$ acting on the cover as a union of spectras of conjugated operators acting on the base space.
We consider the restriction of $\nabla^*\nabla$ acting on the space of sections $s$ of $E_N$ satisfying the automorphic condition
$ s(\gamma.)=e^{\frac{2\pi i p}{N}}s $.
By the exact same conjugation argument, $\nabla^*\nabla$ acting on such sections is isospectral to
$$ \Delta_{N,p}=\nabla^*\nabla-\frac{2i\pi p \nabla \psi}{N}.\nabla- \frac{4\pi^2 p^2}{N^2} \vert \nabla\psi\vert^2_g + \frac{2i\pi p}{N} c ,$$
where $\nabla^*\nabla \psi=c\in C^\infty_{\text{per}}(M_\infty)$ therefore $c$ is bounded on $M_\infty$ and periodic.
Introduce the twisted Laplacian
$$ \Delta_{\theta}=\nabla^*\nabla-2i\theta \nabla \psi.\nabla, -\theta^2 \vert \nabla\psi\vert^2_g + i\theta c,$$
which has periodic coefficients, hence it induces an operator acting on $C^\infty(E)$. 
As above, we need to study the lowest eigenvalue $\lambda_0(\theta)$ of the above operator as function of the angle $\theta$.
We need to decide for what values of $\theta$ this eigenvalue $\lambda_0(\theta)$ vanishes, what is the multiplicity of $\lambda_0(\theta)=0$ and whether the dependence is analytic in $\theta$ near local minimas.

Recall that there is a distinguished element $\gamma $ generating the action on $M_\infty$ by deck transformations. For any closed path $\gamma\in C^\infty(\mathbb{S}^1,M)$ representing the above group element (we make an abuse of notations to use the same letter for the smooth path as well as the element of $\pi_1(M)$ ), 
the parallel transport operator (for the flat connection $\nabla$) along
$\gamma$ is given by the matrix $U=\mathcal{P}_\gamma$ by construction of $E$. 
The idea is to diagonalize $U=\mathcal{P}_\gamma$ over the complex numbers, since the monodromy is unitary, the eigenvalues of 
$\mathcal{P}_\gamma$ will all lie on the unit circle. 
The data of the eigenvalues does not depend on the choice of basepoint on $\gamma\in C^\infty(\mathbb{S}^1,M)$.
We have the following~:
\begin{lemm}
Assume that all eigenvalues of $(\lambda_1,\dots,\lambda_n)\in \{\vert z\vert=1\}\subset \mathbb{C}$ of $\mathcal{P}_\gamma$ are simple. Denote by $\lambda_0(\theta)$ the lowest eigenvalue of the
operator $\Delta_{\theta} $.

 Then
$\lambda_0(\theta)=0$ if and only if $e^{i\theta}$ is an eigenvalue of the monodromy  $\mathcal{P}_\gamma$ of the flat connection $\nabla$.
\end{lemm} 
\begin{proof}
Any section $s\in C^\infty(E_\infty)$ in the kernel of $\Delta_\theta$ is a flat section of the perturbed connection $\nabla + i\theta\alpha\otimes Id$ by repeating exactly the proof of Lemma~\ref{lemm:minimaslambda0} replacing the de Rham differential $d$ with our flat connection $\nabla$.
\end{proof}

Intuitively, the eigenvalues of the monodromy of $\nabla $ all lie on the unit circle and when $e^{i\theta}$ hits such eigenvalue, the twisted Laplacian $\Delta_\theta$ has non trivial kernel, $\lambda_0(\theta)=0$ with multiplicity one.
To repeat the previous argument, we need the spectrum of the monodromy to contain only simple eigenvalues so that the argument using the Kato--Rellich Theorem still holds. 
The function $\theta\in [0,2\pi]\lto \lambda_0(\theta)$ has $n$ local minimas,
when $\theta_1,\dots,\theta_n$ are such that $\lambda_i=e^{i\theta_i}$ where
$(\lambda_1,\dots,\lambda_n)\in \{\vert z\vert=1\}\subset \mathbb{C}$ are the monodromy eigenvalues, $\theta \lto \lambda_0(\theta)$ is locally analytic near each of these minimas and non constant.

Finally, one crucial ingredient we need is some version of the Gaussian bound on the heat kernel of twisted Laplacians, in the vector bundle case. An elegant proof of such bound relies on
the diamagnetic inequalities for the heat kernel whose proof follows from
the Feynman-Kac--It\^o formula. In the comprehensive book by G\"uneysu~\cite[Thm VII.8 p.~111]{Guneysubook}, this is called the Kato-Simon inequality or also Hess–Schrader–Uhlenbrock inequality~\cite{HSU}.

\begin{lemm}[Gaussian bounds twisted case]
Let $(M,g)$ be any smooth geodesically complete Riemannian manifold and $(E,\nabla)\lto M$ a Hermitian bundle over $M$ endowed with some unitary connection $\nabla$. We denote by $\nabla^*\nabla$  the Bochner Laplacian and $\Delta_g$ the scalar Laplacian. 
Then we have the following diamagnetic inequality for heat kernels, for all $(x,y)\in M^2$~:
\begin{eqnarray*}
\boxed{\vert e^{-t\nabla^*\nabla}(x,y) \vert_{E_x\lto E_y }\leqslant e^{-t\Delta}(x,y)}  
\end{eqnarray*}
where the scalar heat kernel dominates the bundle valued heat kernel.

In the particular case where $M$ has Ricci curvature bounded from below, this implies that the heat kernel of the Bochner Laplacian admits Gaussian bounds of Li-Yau type of the form~:
\begin{align*}
\boxed{\vert e^{-t\nabla^*\nabla}(x,y) \vert_{E_x\lto E_y }\leqslant  C_1t^{-\frac{d}{2}  }   e^{-C_2 \frac{\mathbf{d}^2(x,y)}{t} }  }   
\end{align*}
for $C_1,C_2$ that do not depend on $(t,x,y)$.
\end{lemm}

We apply the above Theorem to $M_\infty$ which is a cover of some compact Riemannian manifold therefore, it has therefore Ricci curvature bounded from below and is thus both geodesically and stochastically complete since Brownian motion has indefinite lifetime on $M_\infty$~\cite[Prop XIV.29 p.~192 and Thm XIV.31 p.~193]{Guneysubook}. One could also invoke results of Grigoryan \cite[section 11.4 p.~303]{Grigoryan} which says that manifolds with bounded geometry are always stochastically complete.
The diamagnetic inequality holds true for $e^{-t\nabla^*\nabla}$ by 
~\cite[see eq (2) of Thm 1.1 p.~1998]{Guneysu}.

Then the proof generalizes with minor changes to the bundle setting, we just need to be careful and cut the sum into $n+1$ parts where $n$ pieces deal with eigenvalues accumulating near each minimas, finally this yields:

\begin{thm}
In the geometrical setting of present paragraph, assume that the eigenvalues of the monodromy operator $\mathcal{P}_\gamma$ are all \textbf{simple}. Denote by $\Delta_N$ (resp $\Delta_\infty$) the Bochner Laplacians induced on $E_N:=\pi_N^*E\lto M_N$ (resp $\pi_\infty^*E\lto M_\infty$). Then we have the asymptotic behaviour:  
\begin{align*}
&\lim_{N\rightarrow +\infty}\frac{1}{N}\log\text{det}_\zeta(\Delta_N |_{E_N})=-\frac{d}{ds}|_{s=0} \frac{1}{\Gamma(s)}\int_0^1 \int_{\Omega} Tr_{E_x}\left(e^{-t\Delta_\infty}(x,x)\right)\,\dd V_g(x)\, t^{s-1}\,\dd t\\
&+\int_1^\infty \left(\int_{x\in\Omega} Tr_{E_x}\left(e^{-t\Delta_{\infty}}(x,x)\right)\,\dd V_g(x)\,\right)\frac{\,\dd t}{t}.
\end{align*}
\end{thm}

\end{document}